# Observation of Rydberg moiré excitons

Qianying Hu [1,2,3]†, Zhen Zhan [4]†, Huiying Cui[1,2], Yalei Zhang[4], Feng Jin[1], Xuan Zhao[1,2], Mingjie Zhang[1,2], Zhichuan Wang[1,2], Qingming Zhang[1], Kenji Watanabe[5], Takashi Taniguchi[6], Xuewei Cao[3], Wu-Ming Liu[1,2], Fengcheng Wu[4,7], Shengjun Yuan[4,7]*, Yang Xu[1,2]*

**Affiliations:**

[1] Beijing National Laboratory for Condensed Matter Physics, Institute of Physics, Chinese Academy of Sciences; Beijing, 100190, China

[2] School of Physical Sciences, University of Chinese Academy of Sciences; Beijing, 100190, China

[3] School of Physics, Nankai University; Tianjin, 300071, China

[4] School of Physics and Technology; Wuhan University, Wuhan, China

[5] Research Center for Functional Materials, National Institute for Materials Science; Tsukuba, 305-0044, Japan

[6] International Center for Materials Nanoarchitectonics, National Institute for Materials Science; Tsukuba, Japan

[7] Wuhan Institute of Quantum Technology; Wuhan, China

† These authors contributed equally to this work

*Corresponding author. Email: s.yuan@whu.edu.cn (S. Y.); yang.xu@iphy.ac.cn (Y. X.);

**Abstract:**

**Rydberg excitons, the solid-state counterparts of Rydberg atoms, have sparked considerable interest in harnessing their quantum application potentials, whereas a major challenge is realizing their spatial confinement and manipulation. Lately, the rise of two-dimensional moiré superlattices with highly tunable periodic potentials provides a possible pathway. Here, we experimentally demonstrate this capability through the observation of Rydberg moiré excitons ($X_{RM}$), which are moiré trapped Rydberg excitons in monolayer semiconductor $WSe_2$ adjacent to twisted bilayer graphene. In the strong coupling regime, the $X_{RM}$ manifest as multiple energy splittings, pronounced redshift, and narrowed linewidth in the reflectance spectra, highlighting their charge-transfer character where electron-hole separation is enforced by the strongly asymmetric interlayer Coulomb interactions. Our findings pave the way for pursuing novel physics and quantum technology exploitation based on the excitonic Rydberg states.**

**Main Text:**

The Rydberg states of matter are ubiquitously encountered in various physical platforms, ranging from atoms, molecules to solids (*1–3*). They share common features as exemplified by Bohr's description of the highly excited hydrogen atoms. The large spatial extent of the Rydberg-state wave function promotes large dipole moments with significantly enhanced sensitivity to weak external fields. Over the last two decades, the Rydberg atoms have drawn much more attention owing to the experimental developments in trapping and manipulation of cold atoms, facilitating the study of quantum many-body physics and quantum information processing (*4–8*). Similarly, as the high-order Coulomb bound states of electron-hole pairs emerged in semiconductors, the Rydberg excitons have also been proposed to host various potential applications such as simulating the topological Haldane phase and particular mathematical problems (*3, 9, 10*). Their solid-state nature allows for compatibility with modern semiconductor technologies. However, the requisite controllability on spatial trapping for the Rydberg excitons can be hardly achieved in bulk materials. In this work, we instead utilize the two-dimensional (2D) semiconductor monolayers ($WSe_2$ here), which possess strong light-matter interaction and support excitonic Rydberg states to high orders (*11–16*).

In recent years, the Rydberg sensing technique has appeared in detecting the nearby exotic electronic states and phase transitions by the environmentally sensitive Rydberg excitons in the atomically thin semiconductors (*17–19*). In our experiment, we place 2D moiré superlattices (TBG here, lower layer in Fig. 1) beneath the monolayer $WSe_2$ (upper layer in Fig. 1) to provide spatially periodic modulations. When the wavelength λ of the potential landscape created by TBG is smaller than (or only comparable to) the exciton size $r_B$ [~ 7 nm for the 2s states (*14*)], the Wannier-type exciton's wavepacket is spread over a few moiré unit cells and does not lose its mobile character, as illustrated in the left part of Fig. 1. The optical response of the system is dominated by the Rydberg sensing scheme (*17, 20*).

To realize efficient trapping, the moiré potential must have a spatial profile ($\lambda \propto \frac{1}{\theta}$ at small twist angles $\theta$) that is properly larger than the exciton size (illustrated in the right part of Fig. 1), as has been shown for the ground-state excitons in transition metal dichalcogenide (TMD) heterobilayers (*21–28*). In our system, the accumulated charges in the TBG AA sites could strongly attract the oppositely charged electron or hole of the loosely bound 2s exciton in $WSe_2$, thus achieving spatial confinements of Rydberg moiré excitons ($X_{RM}$) with in-situ controllable interaction strength up to 75 meV tuned by the charge density. The potential wells generated by the periodic charge distribution of TBG in the strong coupling regime ($\frac{\lambda}{r_B} > \sim 2.4$) render an analogue to the optical lattices created by a standing-wave laser beam or arrays of optical tweezers for Rydberg atom trapping (*1, 2, 4*). Due to the nanoscale size ($<<r_B$) of the confining potential well, we would further elaborate that the rigid quantum impurity approximation (*29–36*) is no longer valid for the Rydberg excitons. The $X_{RM}$ realizes electron-hole separation and exhibits the character of long-lived charge-transfer excitons.

**Rydberg sensing**

Figure 2**A** illustrates the typical device schematic. The directly contacted TBG and monolayer WSe$_2$ are encapsulated by hexagonal boron nitride (hBN) dielectrics and graphite electrodes where the gate voltages are applied. Due to the band misalignment between TBG and WSe$_2$, charge carriers are only doped into TBG while WSe$_2$ remains charge neutral within the experimentally accessible gating range. We use a broadband light source to excite the electron-hole pairs in monolayer WSe$_2$ and detect their resonance energies through reflectance contrast ($\Delta R/R_0$) spectroscopy. The energies of excitons carry information about dielectric screenings and interactions with the charges in TBG. More details on device fabrication and optical measurements can be found in Supplementary Materials and Fig. S1.

We first examine the device with relatively large twist angles, i.e., small $\lambda/r_B$. The doping-dependent reflectance contrast ($\Delta R/R_0$) spectrum of device D1 with $\theta = 10°$ TBG ($\lambda/r_B \approx 0.2$ for the 2s excitons) is shown in Fig. 2B. In contrast to the barely changed 1s exciton (the excitonic ground state of WSe$_2$ near 1.71 eV), the 2s exciton (resonance near 1.78-1.8 eV) redshifts and merges into the renormalized band edge with increasing carrier densities (*n*). The observation is similar to that seen in the monolayer graphene/WSe$_2$ system (*20*). For TBG with a twist angle as large as 10°, the low-energy band structure maintains the linear dispersion of isolated graphene. In such small $\lambda/r_B$ limits, the neighboring 2D electron gas (10° TBG here) provides a uniform dielectric background to screen the Coulomb interactions in WSe$_2$. The Rydberg exciton energy can be safely expressed as the subtraction of the binding energy from the quasiparticle band gap, both of which get renormalized by increasing the density of states in TBG. The exciton Rydberg states hence become delicate dielectric sensors to probe the dielectric function and electronic compressibility of the neighboring TBG. See refs.(*20*, *37*) for more details.

The Rydberg sensing scheme also works for the near-magic-angle TBG (device D2, $\theta = 1.14°$, $\lambda/r_B \approx 1.9$) as shown in Fig. 2D. The doping-dependent 2s state (near 1.75-1.78 eV) exhibits a symmetric sawtooth pattern around the zero density, with periodic intensity enhancements on both the electron and hole-doped sides, labeled as ν = -4, -3, …, 3, 4, respectively. Among them, the ν=±4 states correspond to the gap openings when the first moiré subbands of small-angle TBG are empty or fully filled as schematically illustrated in the right panel of Fig. 2E. The features at full filling densities (denoted as $n_s$) allow us to determine the accurate twist angles in a wide range from ~0.73° to 1.6° in our devices (see Supplementary Materials for more details).

On the other hand, the ν = ±3 (±2, ±1, and 0) states, corresponding to 3 (2, 1, and 0) charges per moiré site, are beyond the scope of single-particle band theory. In the right panel of Fig.2D, these features show little temperature dependence in the range of 1.6-10 K, well above the onset of correlated insulating states observed by transport measurements (*38*, *39*). They are consistent with the observations by scanning tunneling microscopy and local electronic compressibility measurements (*40*, *41*), both of which reveal a cascade of four-fold (spin and valley degrees of freedom) symmetry-broken states. We have observed these states with the twist angle ranging from 1.06° to 1.15°, demonstrating the significant role of electronic correlations in the near-magic-angle TBG with band flattening (*38–44*).

In addition to the Rydberg sensing features discussed above, the spatially periodic dielectric screening environment provided by TBG modulates the spectrum as well. The emergence of the replica at higher energies (differ by $\Delta$) is a manifestation of the formation of moiré bands in WSe$_2$ as illustrated in the left panel of Fig. 1E (*20*). New optical transitions between states at high-symmetry points of the mini-Brillouin zone boundary become allowed above the fundamental bandgap, as shown by the schematic in the left panel of Fig.2E (*20*). It is similar to the emergence of new bright resonances due to the Bragg-umklapp scatterings in the exciton dispersion picture (*45*, *46*). The experimentally observed energy separation between the 2s state (or band edge) and its replica $\Delta$ is about 32 meV, slightly larger than the value $h^2/6m_r\lambda^2$=22 meV (where $h$ is Planck's constant and the reduced exciton mass $m_r$ is about 0.15 free electron mass) expected from the weak perturbation limit.

**Rydberg moiré excitons at smaller twist angles**

When the twist angle $\theta$ is further reduced, we observe a significant enhancement of the interlayer Rydberg exciton-charge interactions, indicating access to the strong coupling regime. Similar reflectance spectrum measurements in device D3 with $\theta \approx 0.6°$ TBG ($\lambda \approx 23.5$ nm, $\lambda/r_B \approx 3.6$) is shown in Fig. 3A (color contour plot) and 3B (linecuts). Upon doping the TBG with either positively or negatively charged carriers, the 2s resonance peak at 1.783 eV splits into multiple branches. The most obvious four branches are traced by the dashed curves in Fig. 3B. They have nonmonotonic doping dependences. The resonance energies redshift first and then blueshift, reaching their minima at densities $n_m \approx 3.8 \times 10^{12}$ cm$^{-2}$ for the electron-doped side and $-4.7 \times 10^{12}$ cm$^{-2}$ for the hole-doped side. Some of the main branches could survive at relatively high temperatures up to 140 K, as shown in Fig. S2. The splitting behavior reminds us of the multiple peaks of the ground-state moiré excitons observed in the TMD heterobilayers, suggesting the exciton wavefunction residing at different moiré stacking sites and experiencing inequivalent potentials (20-27). We attribute our observations to the formation of Rydberg moiré excitons (X$_{RM}$).

The lowest-energy branch of the X$_{RM}$ possesses the most prominent doping dependence that is at first sight striking. It approaches the 1s exciton energy, by only ~10 meV larger at $n_m$. The energy shift magnitude |$E_{shift}$| from the charge neutrality point is extracted in Fig. 3C, where the carrier density is normalized by its full filling density $n_s$=8.37×10$^{11}$ cm$^{-2}$. It nearly shows a linear dependence on the density for |$n/n_s$|<~4, as guided by the dashed lines in Fig. 3C. The |$E_{shift}$| reaches its maximum $E_m$=73 meV on the electron-doped side, which is much larger than the internal electron-hole binding energy of the 2s state [~15 meV for the WSe$_2$ in proximity to undoped monolayer graphene (*20*)]. A simple dielectric screening cannot account for such large energy shifts.

The interaction between an exciton and a uniform Fermi sea has been extensively studied (*29–36*). The conventional wisdom of dealing with the exciton is treating it as a rigid and mobile impurity in a degenerate Fermi system. However, such an assumption is no longer valid here owing to the large spatial extension of the Rydberg excitons and the presence of long-wavelength moiré potentials. Because the change in intralayer binding energy is much smaller compared to the interlayer interaction, the energy shift of X$_{RM}$ reflects the Coulomb interaction energy between the 2s exciton and free charges in TBG. A likely scenario is related to the spatially confined charge distributions in TBG that can help promote unequal interlayer interactions for the constituent electron and hole of the exciton.

We hence carry out numerical simulations to extract the real-space charge distribution in TBG at different doping levels as shown in Fig. 3E (see Supplementary Materials and Fig.S3-S5 for further information). Starting from the charge neutral point, the local charge density rises drastically at the AA-stacked regions first while is merely changed at the AB/BA-stacked regions. This is closely related to the much larger local density of state (LDOS) of the AA site at small densities. While the AA-stacked region has a radius of ~2.6 nm (estimated from the full width at half maximum of the spatially enhanced charge accumulation peak) that is much smaller than $r_B$, the areas of the AB/BA stacked region are greatly enlarged due to lattice reconstruction (with the schematic superlattices shown in the lowest map in Fig.3E) (*47*). Taking the electron-doped side as an example, the accumulated charges centered at the AA-stacked region create deep and narrow potential wells for trapping the hole of the exciton, while the trend to minimize the repulsion energy will push the electron of the exciton towards the AB and BA-stacked regions as schematically shown in the inset of Fig.3D. The role of electron and hole is reversed on the hole-doped side. This process renders a spatial separation of the electron-hole pair, supporting a charge-transfer type exciton that typically only forms in molecular crystals with electron/hole occupying adjacent molecules or across an interface between two kinds of materials (*48*).

In this scenario, with one charge residing at the AA site and the other charge mostly on the AB/BA site, the electron and hole of the Rydberg exciton possess highly asymmetric electrical potential energies. The $E_{\text{shift}}$ can then be approximately estimated from the difference in attraction on the AA site and repulsion on the AB/BA site as $E_{\text{shift}} \approx (eU_{AA} - eU_{AB/BA}) \propto (n_{AA} - n_{AB/BA})$, where $U_{AA}$ ($U_{AB/BA}$) and $n_{AA}$ ($n_{AB/BA}$) are the electric potential and charge density at the AA (AB/BA) site, respectively. Quantitatively, the calculated $n_{AA} - n_{AB/BA}$ as a function of $n/n_s$ is presented in Fig. 3D, where a similar nonmonotonic trend as $E_{\text{shift}}$ is observed, with the critical densities ($n_m$) almost identical to our experimental observations on both the electron and hole-doped sides. As the doping reaches $n_m$ where LDOS of TBG nearly equalizes at the two sites, the charge density at the AB/BA site begins to grow faster than that in the AA site (see detailed charge-filling maps in Fig. S4) and the lattice becomes more evenly filled at higher doping densities. It explains the blueshift for $X_{RM}$ at $|n|>|n_m|$ and the convergence of the multiple branches in the large density limit.

Meanwhile, considering the size of 2s exciton ($r_B$~7 nm) that is about an order of magnitude larger than the interlayer spacing $d$ ($\approx$0.5 nm), the interlayer Coulomb interaction with TBG charges can be much stronger than the intralayer one. It is thus possible to see the pronounced energy shift of the lowest-energy branch that is a few times larger than its original binding energy after the TBG is doped. We have performed control experiments by inserting additional hBN spacers to increase the interlayer distance as shown in Fig. S6. The energy shift is then strongly suppressed and the system exhibits characteristics with weak interlayer interactions.

**Crossover between the weak and strong coupling regimes**

To better reveal the evolution to the strong coupling regime, more devices are fabricated and measured with twist angles ranging from ~0.6° to 1.23°, as presented in Fig.4A. As the twist angle becomes smaller, the maximum redshift (denoted as $E_m$, at the critical density $n_m$) of the 2s state or the lowest-energy $X_{RM}$ keeps growing, accompanied by the emergence of more replica states. Notably, shown in the last panel in Fig.4A, the energy separation between 1s and 2s states at $n_m$ nearly vanishes in the smallest-twist-angle device where $\theta$ is slightly smaller than 0.6°. It suggests a strong interlayer Coulomb interaction of the $X_{RM}$ that is even comparable to the binding energy of the 1s state and the possibility of forming new molecular states (*49*).

We summarize the twist angle dependences of $E_m$ and $n_m$ on the electron-doped sides in Fig. 4B and 4C, respectively. The $E_m$ is a direct evaluation of the maximum interlayer interaction, and the experimentally obtained $n_m$ at three twist angles is in good agreement with that obtained from the theoretical calculations given in Fig. S5. The $n_m/n_s$ and $E_m$ share a similar trend upon varying the angle $\theta$, likely stemming from the positive correlation between charge accumulation at each moiré site and the interlayer Coulomb interaction energy. Meanwhile, an obvious reduction of $X_{RM}$ linewidth (extracted for the lowest-energy branch) from ~8 meV to ~1.5 meV is observed with decreasing the twist angle into the strong coupling regime (Fig. 4D). It is consistent with our interpretation of $X_{RM}$ where the spatially separated electron-hole configuration is likely to support longer lifetimes.

To summarize, we develop and experimentally demonstrate a new method in spatially confining and manipulating the Rydberg excitons by the long-wavelength moiré potentials. The strongly bound $X_{RM}$ complex can be dominated by interlayer interactions and approach the energy of ground-state excitons. The system provides easy access to control the potential well depth by electrostatic doping, to tune the moiré wavelength by the twist angle, and to achieve longer lifetimes guaranteed by the electron-hole separation. All these features would be beneficial for further realizing excitonic Rydberg-Rydberg interactions and coherent controls. Our study could open up unprecedented opportunities for the implementation of quantum information processing and quantum simulation based on the versatile Rydberg states in solid-state systems (*9*, *10*).

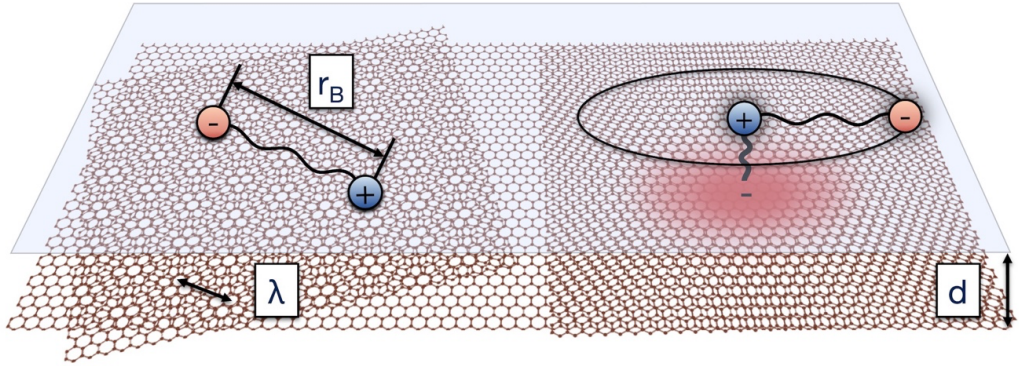

**Fig. 1. Schematic illustration of the interplay between a Rydberg exciton (size $r_B$) and moiré superlattices (TBG here) with small and large periodicity (wavelength $\lambda$).** In the small $\lambda/r_B$ limit (left panel), the moiré system provides a nearly uniform dielectric environment and the optical response of the exciton is dominated by the Rydberg sensing features. The exciton maintains its mobile character. In the large $\lambda/r_B$ limit (right panel), the Rydberg exciton can be confined by the moiré potential well generated by the accumulated charges in the AA site of TBG. The $r_B$ is typically an order of magnitude larger than the interlayer spacing $d$, rendering much stronger interlayer interaction than the intralayer one in the formed Rydberg moiré exciton ($X_{RM}$) complex.

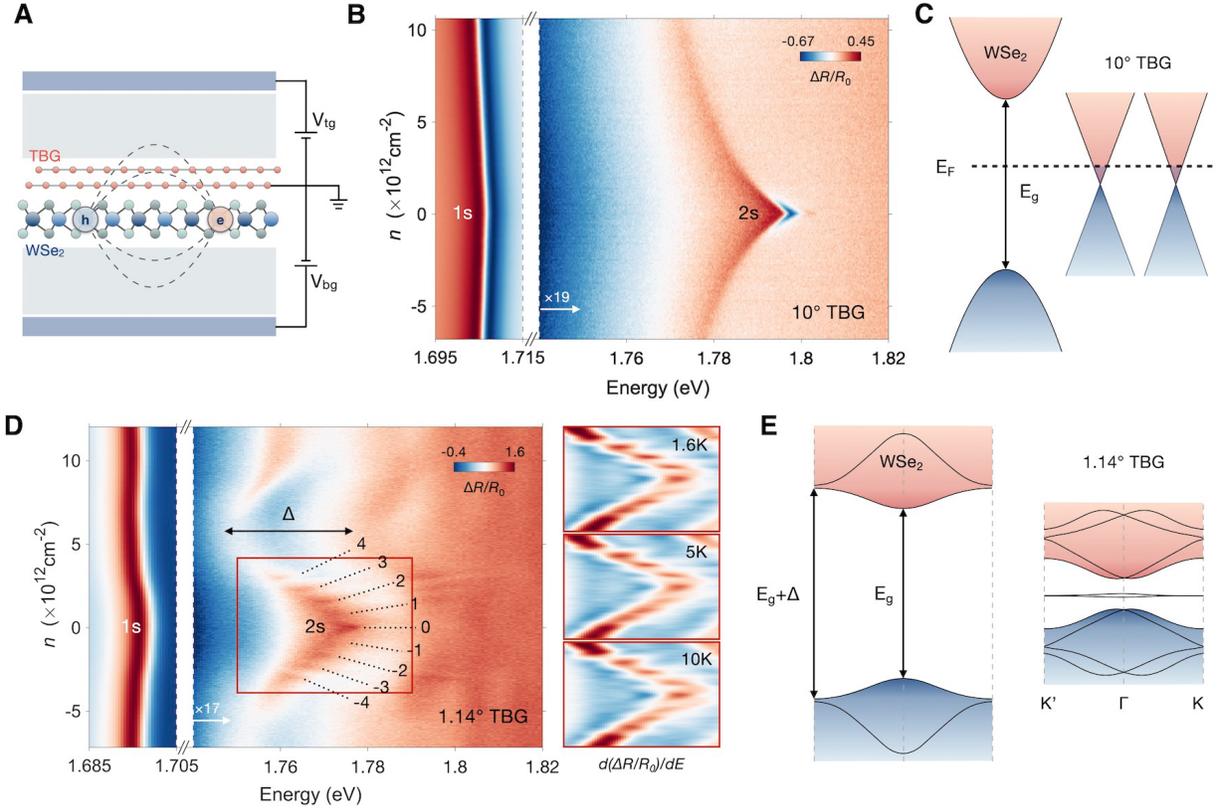

**Fig. 2. Rydberg sensing of WSe$_2$ adjacent to 10° and 1.14° TBG.** (**A**) Schematic structure of a typical device with electrically grounded TBG and monolayer WSe$_2$ embedded in hBN/graphite dual gates. (**B**) Doping-dependent reflectance contrast ($\Delta R/R_0$) spectrum of device D1 with 10° TBG ($\lambda/r_B$=0.2). The WSe$_2$ serves as a dielectric sensor, whose 2s exciton energy reflects the dielectric screening of the neighboring TBG. (**C**) Schematic band alignment between monolayer WSe$_2$ and 10° TBG. (**D**) Doping-dependent $\Delta R/R_0$ spectrum of device D2 with 1.14° TBG ($\lambda/r_B$=1.9). Sawtooth features appear at filling factors -4 to 4 (highlighted by the red rectangle), and a new replica is observed at higher energies of $\Delta$=32 meV. The right three panels show the temperature-insensitive sawtooth features at 1.6 K, 5 K, and 10 K, respectively. (**E**) Schematic band structure of near-magic-angle TBG and the adjacent monolayer WSe$_2$. The TBG features gap openings and flat bands, which result in the band insulating states ($\nu = \pm 4$) and the cascade of symmetry-breaking phase transitions ($\nu = 0, \pm 1, \pm 2, \pm 3$) observed in (D), respectively. Meanwhile, the spatially periodic screening of TBG folds the band of WSe$_2$ into the mini-Brillouin zone, generating a new optically-allowed transition at $E_g+\Delta$. The $\Delta R/R_0$ spectra above 1.74 eV in (B) and (D) are multiplied by a factor of 19 and 17 for better illustration, respectively.

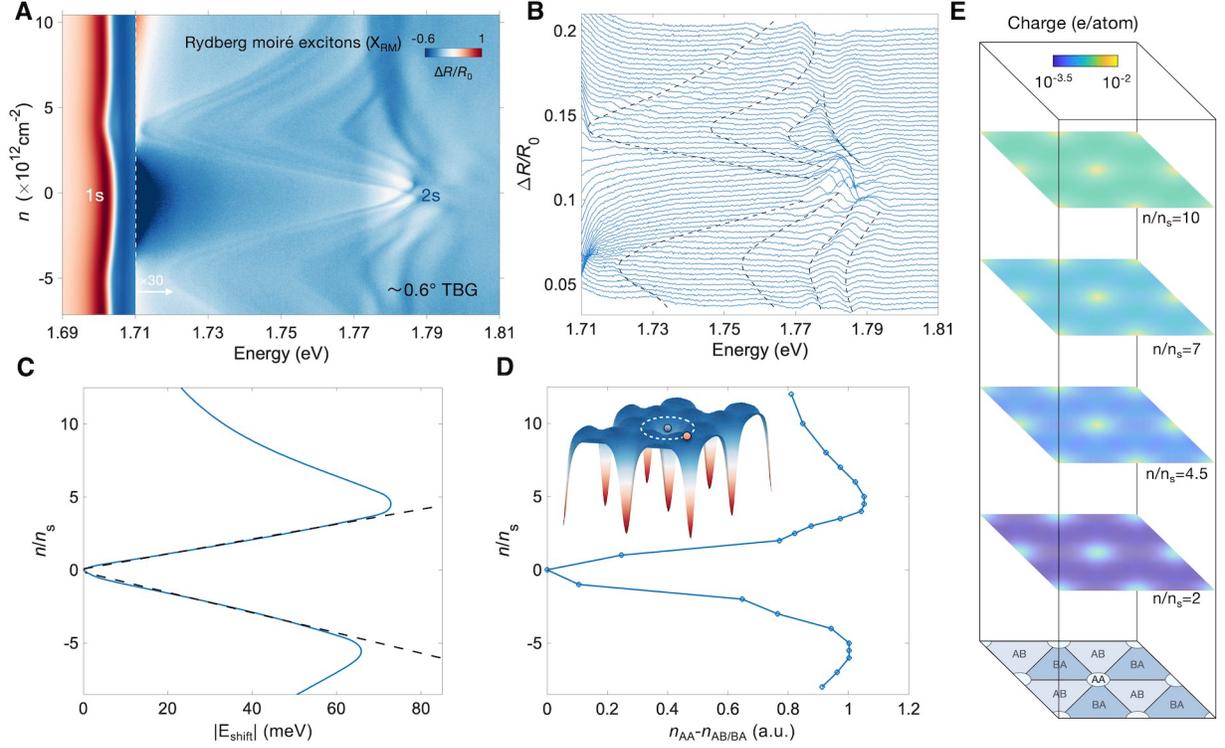

**Fig. 3. Rydberg moiré exciton ($X_{RM}$) formation in WSe$_2$ adjacent to 0.6° TBG.** (**A**) Reflectance contrast spectrum of device D3 with 0.6° TBG ($\lambda/r_B$=3.6). The WSe$_2$ 2s exciton resonance splits into multiple branches and exhibits nonmonotonic dependences when the TBG is doped. The $\Delta R/R_0$ spectra above 1.71 eV are multiplied by 30. (**B**) Linecuts of (A) showing the $X_{RM}$ evolution with tuning densities. Curves are shifted vertically for clarity. Dashed curves are guides to eyes tracing four main branches of $X_{RM}$ on electron- and hole-doped sides. (**C**) Extracted energy shift of the lowest-energy branch in (B) as a function of $n/n_s$ ($n_s$ denoting the full filling density of the first moiré band). (**D**) Calculated local charge density difference between the AA and AB/BA sites with varying $n/n_s$. Insets are the schematic exemplification of the lowest-energy $X_{RM}$ confinement on the electron-doped side. The moiré potential landscape facilitates the charge-transfer type exciton configuration with the hole (blue sphere) residing on the AA site and electron (red sphere) on the AB/BA site. The $n_{AA} - n_{AB/BA}$ versus $n/n_s$ approximately reproduces the energy shift in (C) as $E_{\text{shift}} \approx (eU_{AA} - eU_{AB/BA}) \propto (n_{AA} - n_{AB/BA})$. (**E**) Calculated spatial charge distribution (in logarithmic scales) of 0.6° TBG at representative doping densities. The lowest map is a schematic of relaxed TBG moiré superlattices with AA, AB, and BA sites marked.

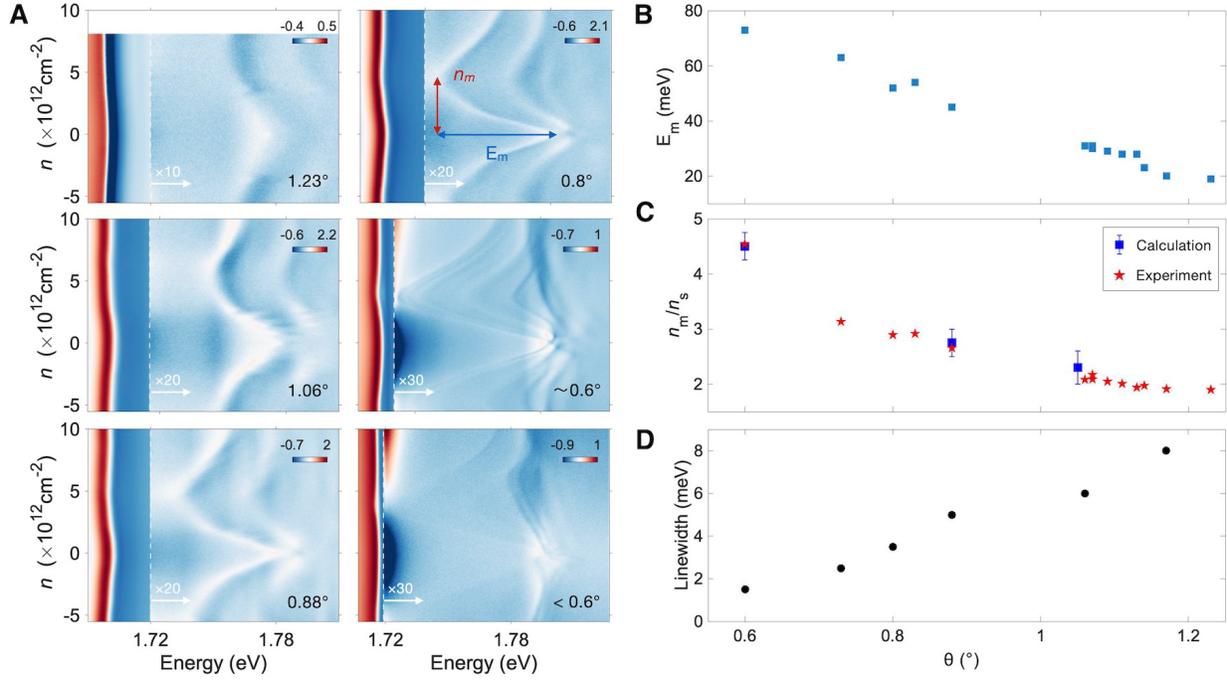

**Fig. 4. Twist angle dependences and crossover to the strong-coupling regime.** (**A**) Doping-dependent reflectance contrast spectra of devices with 1.23°, 1.06°, 0.88°, 0.8°, 0.6°, and <0.6° TBG. The nearly parallel 2s resonance and replica evolve into $X_{RM}$ with increasing $\lambda/r_B$ (decreasing twist angle). (**B** to **D**) Twist angle dependence of the maximum energy shift $E_m$ (B), the normalized critical density $n_m/n_s$ (C), and the estimated linewidth (D) extracted from the 2s resonance or the lowest-energy branch of $X_{RM}$ on the electron-doped side. The critical density expected from the calculated $n_{AA}$-$n_{AB/BA}$ is marked in blue (C), in good agreement with the experimental results. The strongly reduced optical linewidth for θ<~0.9° is an indication of longer lifetimes of the $X_{RM}$, in accordance with its charge-transfer character in the strong coupling regime.

# Supplementary Materials

**Materials and Methods**

Device fabrication and electrostatic gating

The device schematic and fabrication process of a typical device are shown in Fig. S1. The WSe$_2$, hBN, graphene, and few-layer graphite are first mechanically exfoliated from bulk crystals onto SiO$_2$/Si substrates (285 nm oxidation layer). Flakes with desired thickness and size are then selected according to their optical contrast. The flakes are picked up layer by layer using the dry-transfer and tear-and-stack methods (*50–52*). The stack is then released onto SiO$_2$/Si substrates with pre-patterned gold electrodes.

Optical measurements

The devices are loaded into a close-cycle cryostat attoDry2100 where the reflection contrast measurements are performed at 1.7 K unless otherwise specified (Fig. 2B and the first panel of Fig.4A are obtained in Montana Cryostation s50 at T = 15 K). A halogen lamp is employed as the white light source, whose output is first collected by a single-mode fiber and then collimated by a ×10 objective lens. A low-temperature compatible objective (NA=0.82) focuses the beam onto the sample. The beam diameter is about 1 μm and the power is lower than 1 nW. The reflected light from the sample is collected by the same objective and detected by a spectrometer. The reflectance contrast ($\Delta R/R_0$) spectrum is obtained by comparing the reflected light spectrum from the sample ($R$) with that from the substrate immediately next to the sample ($R_0$) as $\Delta R/R_0 = (R - R_0)/R_0$. Keithley 2400 source meters are employed to apply the gate voltages. In small-angle twisted devices, the lowest-energy branch of the Rydberg moiré exciton can fall into the tail of the 1s resonance and become hard to identify. We average the spectrum at all gate voltages and have it subtracted from the original spectrum to remove the unwanted background.

Calibration of carrier density.

Carrier density induced by the gate voltage in device D2 (with near-magic twist angle) is calibrated by the spectral feature of Landau levels (LLs) in the monolayer graphene region under perpendicular magnetic fields ($B$). The carrier density of filling one degeneracy lifted LL is determined by $n_0=eB/h$, where $e$ is the elementary charge. From the corresponding gate voltage span $V_0$, we can extract the geometrical capacitance per unit area as $C_g = en_0/V_0$ accurately. The carrier density induced by gate voltage $V_g$ is then obtained by $n = n_0 V_g / V_0$. Carrier densities in other devices are evaluated by the capacitances calculated from the hBN thicknesses measured by atomic force microscopy.

Determination of the twist angle.

Twist angles of small-angle TBG are determined by the spectral feature of the full-filling gap at superlattice filling factor $\nu = \pm 4$. The gap manifests as a horizontal line with an abrupt change in the intensity of the spectra for regions with twist angles ranging from ~0.9° to 1.6° (see examples in Fig.4A). When the twist angle becomes smaller than ~0.8°, the jump in intensity could no longer be recognized. Instead, the sudden change of the $|E_{\text{shift}}|$ slope of the lowest-energy X$_{\text{RM}}$ branch becomes an identifiable signature. After extracting the corresponding carrier density $n_s$ at the full fillings, the twist angle could be obtained through $n_s \approx 8\theta^2/\sqrt{3}a^2$ ($a = 0.246$ nm denoting the graphene lattice constant). For the regions with twist angles smaller than 0.7°, however, no identifiable features could be found in the spectra at full fillings, suggesting no gap opening at such small twist angles (in agreement with our calculation of the density of states shown in Fig. S3). The twist angle of ~0.6° for device D3 is estimated through the nearly linear angle dependence of $E_m$ (Fig. 4B) and then verified by its $n_m$ value as compared with the theoretical calculation (Fig. 4C).

**Supplementary Text**

Band structure, density of states and charge distribution calculations.

The twisted bilayer graphene (TBG) is constructed by identifying a common periodicity between the two layers, which has a moiré length $\lambda = a/(2\sin(\theta/2))$ (*53*). Then, the system is fully relaxed (both in-plane and out-of-plane) via the semi-classical simulation package LAMMPS (*54*). The intralayer and interlayer interactions of the graphene layers are simulated with long-range carbon bond-order potential (*55*) and Kolmogorov-Crespi potential (*56*), respectively. We use an atomistic tight-binding (TB) model to calculate the band structure, density of states (DOS) and charge distributions of the TBG. The TB Hamiltonian for TBG, considering only $p_z$ orbitals of carbon atoms, is

$$H = \sum_i \epsilon_i |i\rangle\langle i| + \sum_{\langle i,j \rangle} t_{ij} |i\rangle\langle j|$$

where $|i\rangle$ is the $p_z$ orbital located at $\mathbf{r}_i$, $\epsilon_i$ is the on-site energy of orbital $i$, $\langle i,j \rangle$ is the sum over index $i$ and $j$ with $i \neq j$, $t_{ij}$ is the hopping integral between sites $i$ and $j$

$$t_{ij} = m^2 V_{pp\sigma}(|\mathbf{r}_{ij}|) + (1 - m^2)V_{pp\pi}(|\mathbf{r}_{ij}|)$$

Here, $\mathbf{r}_{ij} = \mathbf{r}_j - \mathbf{r}_i$, $m$ is the direction cosine along $\mathbf{e}_z$ that is perpendicular to the graphene layer. The Slater-Koster (SK) parameters $V_{pp\pi}$ and $V_{pp\sigma}$ is

$$V_{pp\pi}(|\mathbf{r}_{ij}|) = -\gamma_0 e^{q_\pi\left(1-\frac{|\mathbf{r}_{ij}|}{d}\right)} F_c(|\mathbf{r}_{ij}|)$$

$$V_{pp\sigma}(|\mathbf{r}_{ij}|) = \gamma_1 e^{q_\sigma\left(1-\frac{|\mathbf{r}_{ij}|}{h}\right)} F_c(|\mathbf{r}_{ij}|)$$

where $b = 0.142$ nm and $c = 0.335$ nm are the nearest in-plane and out-of-plane distances, respectively. The SK parameters are $\gamma_0 = 2.8$ eV and $\gamma_1 = 0.44$ eV. The parameters $q_\sigma$ and $q_\pi$ satisfy $q_\sigma/c = q_\pi/b = 0.2218^{-1}$, and the smooth function is $F_c(r) = (1 + e^{(r-r_c)/l_c})^{-1}$ being $l_c$ and $r_c$ chosen to be 0.0265 and 0.5 nm, respectively.

The band structure of the TBG is obtained by direct diagonalization of the TB Hamiltonian. The DOS and charge distribution are calculated by using the tight-binding propagation method (TBPM) implemented in the TBPLaS package (*57*), and the results for three representative twist angles are given in Fig. S3. In TBPM, a random state in the real-space $|\varphi_0\rangle = \sum_i a_i |i\rangle$ is used as an initial state, and the coefficients $\{a_i\}$ are normalized random complex numbers. The DOS is calculated from (*58*)

$$D(E) = \frac{1}{2\pi} \int_{-\infty}^{\infty} e^{iEt} \langle\varphi_0|e^{-iHt/\hbar}|\varphi_0\rangle dt,$$

and the charge density at site $i$ is given by (*59*)

$$n_i = \frac{N}{2\pi} \int_{-\infty}^{\infty} \left|\langle i|e^{-iHt/\hbar}\sqrt{f(H)}|\varphi_0\rangle\right|^2 dt,$$

where $N$ is the total number of electrons, $f(H) = 1/(1 + e^{(H-\mu)/k_B T})$ is the Fermi-Dirac distribution operator, $\mu$ is the electronic chemical potential, $k_B$ is the Boltzmann constant and $T$ is the temperature. The local charge density $n_{AA/AB}$ at the high-symmetry stacking regions AA (AB) in Fig. 3 and Fig. S4-S5 are obtained from

$$n_{AA/AB} = \sum_{|\mathbf{r}_i - \mathbf{r}_{AA/AB}| < r_0} n_i,$$

where $\mathbf{r}_{AA/AB}$ is the center of the AA (AB) region and $r_0$ is set as 2 nm (smaller than the approximated size of AA/AB regions). For more details of the numerical methods, we refer to Refs. (*57–59*).

Role of the interlayer distance between WSe$_2$ and TBG

For the loosely-bound Rydberg exciton, the interlayer distance $d\sim 0.5$ nm is much smaller than its intralayer distance $r_B \sim 7$ nm, resulting in a stronger Coulomb attraction with TBG charges that contributes to the pronounced energy shift of X$_{RM}$. It is then possible to tune the interlayer interaction strength by inserting an additional thin hBN spacer to change the spacing between the WSe$_2$ and TBG. The results are shown in Fig. S6. For the $\theta\sim 0.8°$ device in the strong coupling regime, the doping dependences of the Rydberg excitons are obviously weakened with increasing the thickness of the hBN spacer, indicating the suppression of interlayer Coulomb interactions. Notably, the maximum shift $E_m$ of the lowest-energy branch roughly follows a $1/d$ dependence, consistent with the long-range Coulomb nature.

We also extract the maximum energy separation $\Delta_m$ between the two main lowest-energy branches from the devices with directly contacted TBG/WSe$_2$ and with a hBN spacer in Fig. S6D. We plot a dashed curve ($=h^2/6m_r\lambda^2$) to show the expected replica separation value with weak interlayer interactions (*19*). For the directly contacted devices, more deviation from the curve is observed at smaller twist angles. While for the devices with increased interlayer spacings, the experimentally obtained $\Delta_m$ approximately falls on the dashed curve. These observations further support our interpretation of the X$_{RM}$, and emphasize the role of interlayer spacing in the interlayer interaction strength.

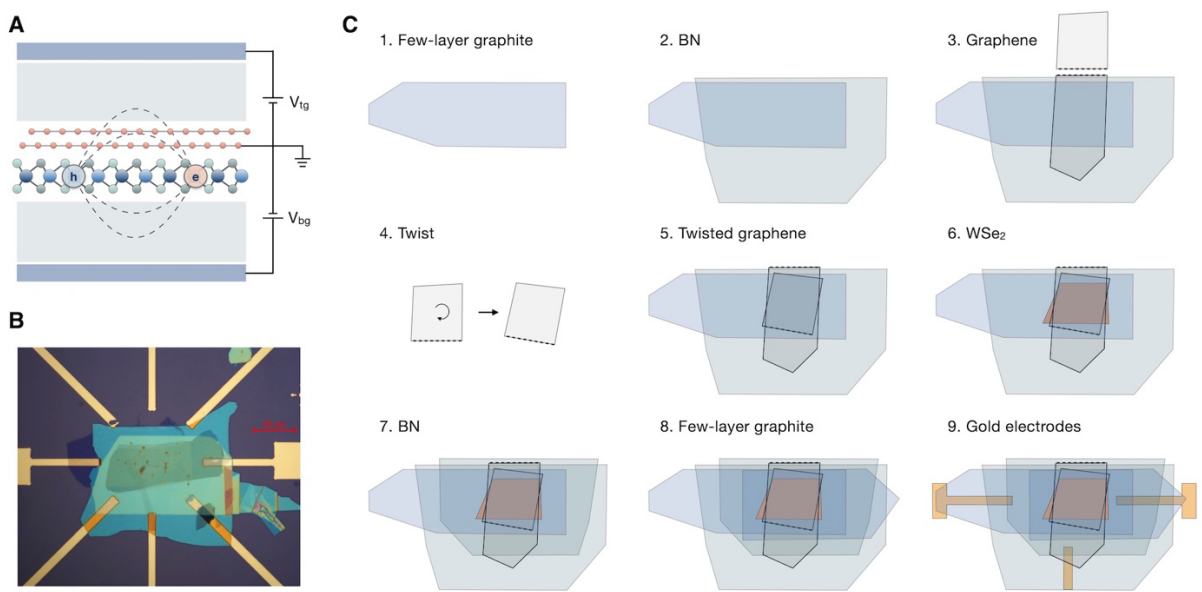

**Fig. S1. Device Fabrication.** (**A** to **C**), Device schematic (A), optical image (B) and fabrication process (C) of a typical device.

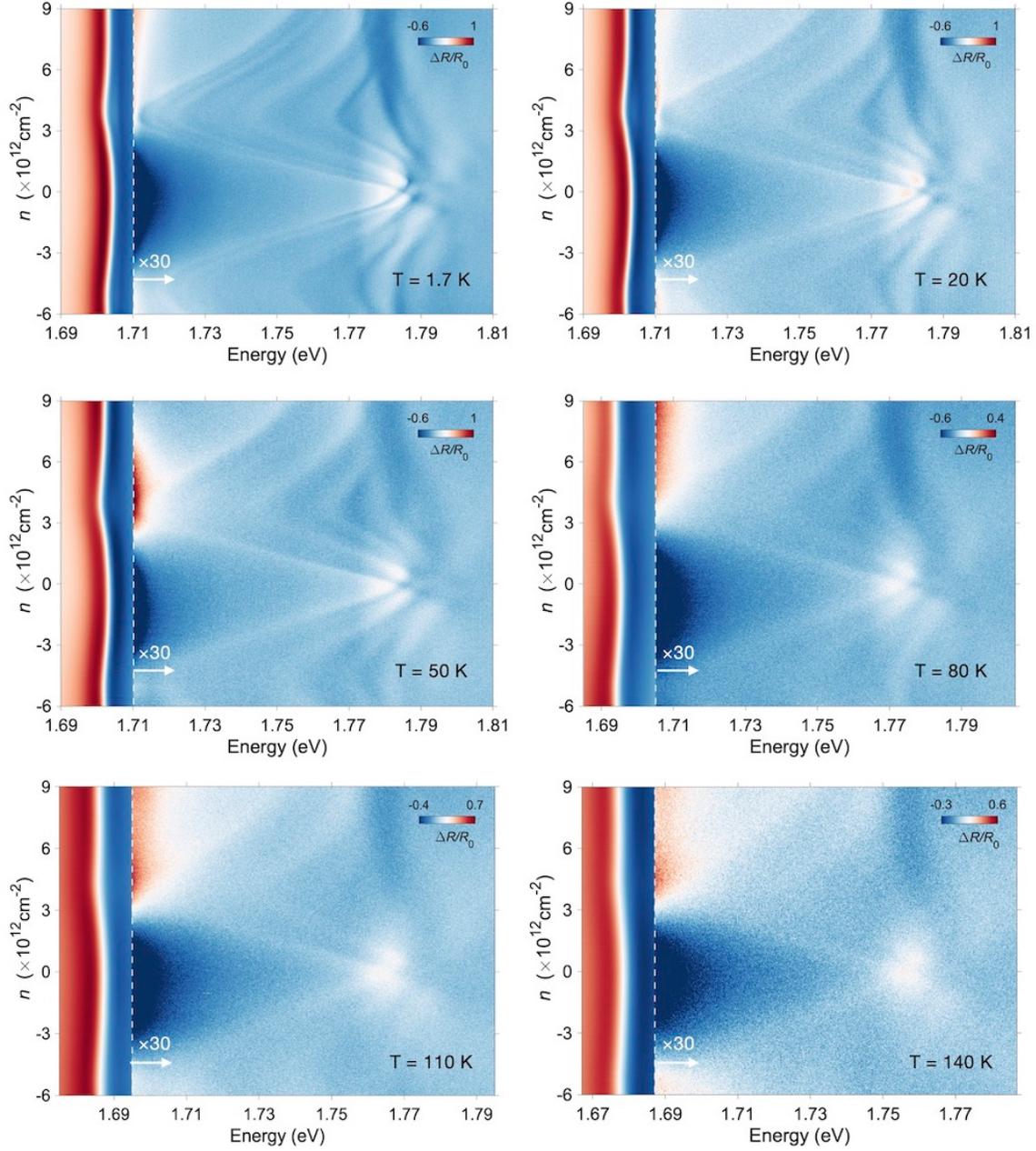

**Fig. S2. Temperature dependence of the Rydberg moiré exciton.** The lowest-energy branch of the $X_{RM}$ survives to a temperature as high as ~140 K, in accordance with the large interlayer binding energies. The spectra for energies above the dashed lines are multiplied by a factor of 30 for clarity.

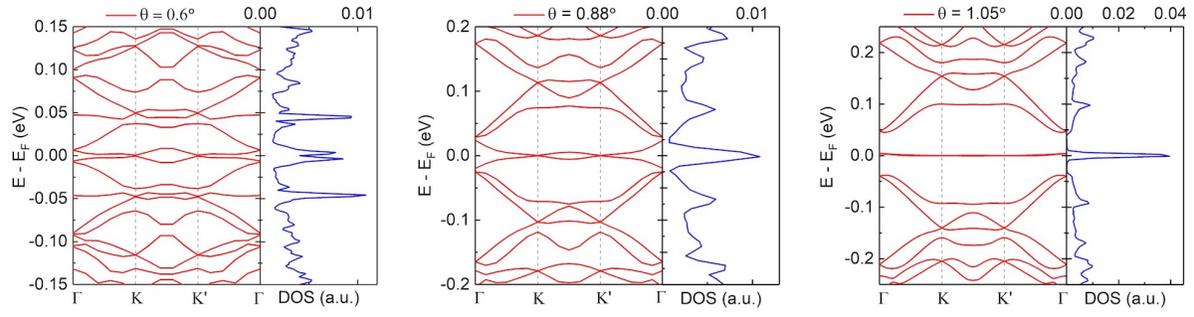

**Fig. S3. Calculated band structure (in red) and density of states (blue) of 0.6°, 0.88°, and 1.05° TBG.** The gap opening at the full filling vanishes at small twist angles (such as 0.6°).

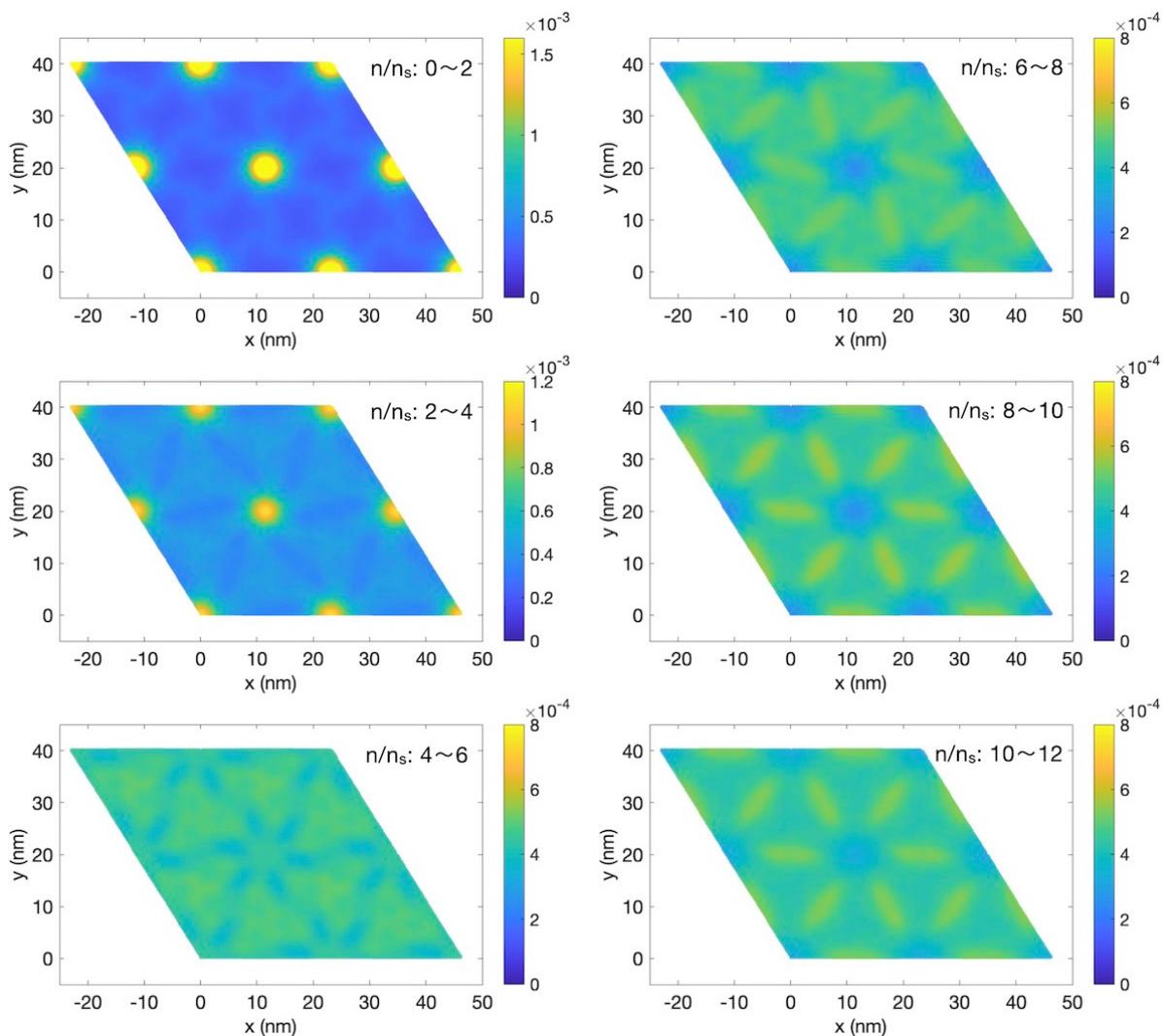

**Fig. S4. Calculated charge-filling map of 0.6° TBG at different doping density ranges.** The maps show the differences between the charge distribution at two values of $n/n_s$, which gives the charge-filling distribution in this range. At low doping densities (e.g. $n/n_s <\sim 4$), the charges prefer to fill onto the AA sites (the regions in bright yellow). As the density becomes higher, the charges fill more onto the AB/BA stacked regions, with the less filled AA sites in dark blue.

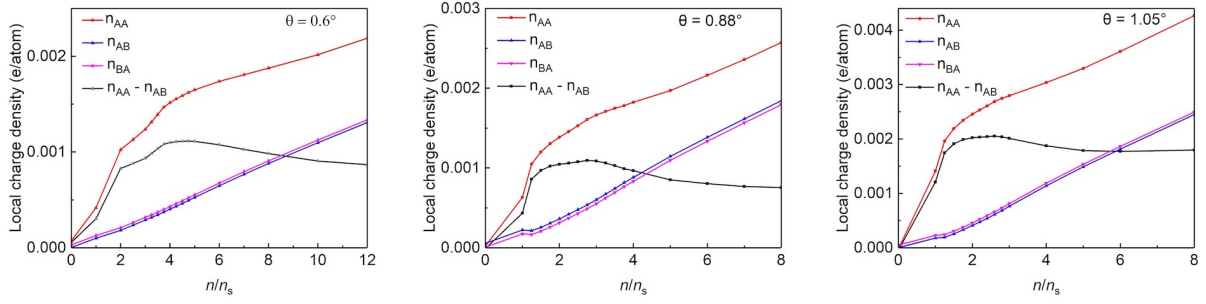

**Fig. S5. Calculated local charge density evolution of 0.6°, 0.88°, and 1.05° TBG.** The local charge densities at AA, AB and BA sites increase with the total carrier density $n$, while the differences between $n_{AA}$ and $n_{AB}$ exhibit non-monotonic trends.

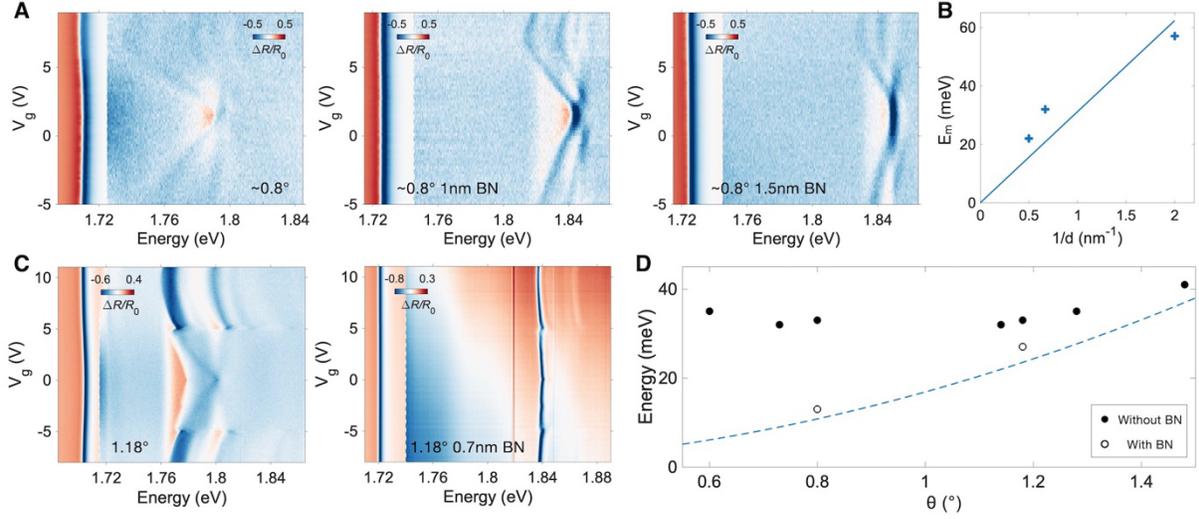

**Fig. S6. Effect of the WSe$_2$-TBG interlayer distance.** (**A**) For a $\theta$~0.8° device with increasing hBN thickness, the X$_{RM}$ evolves into near-parallel resonances with the evidently reduced energy shift upon doping. Spectra with energy above the dashed line are multiplied by a factor for clarity. (**B**) Dependence of the maximum energy shift $E_m$ versus $1/d$ extracted from a, with $d$ denoting the interlayer distance. The solid line is a linear fit. (**C**) The effect of an hBN spacer for devices with relatively weak interlayer interactions ($\theta$=1.18°). The energy shift is reduced after the insertion of the hBN spacer and the energy separation between the 2s and its replica only decreases slightly. (**D**) Extracted energy separation maxima $\Delta_m$ from the devices with (open symbol) or without BN spacer (filled symbol). Data from the weak coupling regime (large twist angle or with a BN spacer) roughly follows the angle dependence expected from the periodic dielectric screening picture (dashed curve), while data beyond the weak coupling regime (small twist angles without BN spacer) show obvious deviations.

24